\documentclass[prd,showpacs,preprintnumbers,amsmath,amssymb,nofootinbib]{revtex4} 
\everymath{\displaystyle} 
\usepackage[dvips]{graphicx} 
\usepackage{longtable}

\def\simge{\mathrel{%
       \rlap{\raise 0.511ex \hbox{$>$}}{\lower 0.511ex \hbox{$\sim$}}}}
\def\simle{\mathrel{
       \rlap{\raise 0.511ex \hbox{$<$}}{\lower 0.511ex \hbox{$\sim$}}}}
\begin{document}

\preprint{SU-4252-816,TKYNT-05-19, IMSc/2005/09/21}

\title
{Semi-superfluid strings in High Density QCD}

\author{A. P. Balachandran$^{1}$
,  S. Digal$^{2,3}$,  and T. Matsuura$^{2}$}

\affiliation{
$^{1}$Physics Department, Syracuse University, Syracuse,
New York 13244-1130, USA\\
$^{2}$Department of Physics, University of Tokyo,
  Tokyo 113-0033, Japan\\
$^{3}$Institute of Mathematical Sciences, C.I.T. Campus, 
Taramani, Chennai  600 113, India}        

\begin{abstract}
We show that topological superfluid strings/vortices with flux tubes
exist in the color-flavor locked (CFL) phase of color superconductors.
Using a Ginzburg-Landau free energy we find the configurations of these 
strings. These strings can form during the transition from the normal 
phase to the CFL phase at the core of very dense stars. We discuss an interesting 
scenario for a network of strings and its evolution at the core of dense 
stars.
\end{abstract}
\maketitle

\section{Introduction}

Color superconductivity is expected to be the ground state for
high baryon chemical potential $\mu$ and low temperatures \cite{BL}. It is
believed that such a state of matter may exist at the core of very dense
stars. To find any signature of its existence, it is important to study
various properties of the color superconducting phase. When all the quarks
are massless, there are superconducting phases, namely 
the color-flavor locked (CFL) and the 2SC phases\cite{BL}. 
In the CFL phase, all quark
flavors take part in the condensation, but in the 2SC phase, one quark does 
not participate in the condensation. With realistic values of quark masses and
charge neutrality conditions, the CFL phase gets modified to phases known
as mCFL, dSC \cite{taeko}, uSC \cite{uSC}, gCFL, 
g2SC \cite{Alford:2003fq,Shovkovy:2003uu},
and FFLO phases \cite{loff,Alford:2000ze} etc. 
One of the interesting properties of 
a color superconductor is that it is also a superfluid. This is
because apart from local symmetries, such as $SU(3)_C$ of color symmetry
and $U(1)_{EM}$ of electromagnetism, certain global symmetries are also
spontaneously broken in the superconducting phase. The global symmetries
spontaneously broken in the superconducting phase are $SU(3)_F$ (the flavor
symmetry) and $U(1)_B$ (the baryon number symmetry). One expects defect
solutions such as flux tubes or vortices to be present in the superconductor
since it is both a superconductor and a superfluid. Recently there have been
lots of studies of superfluid vortices \cite{ZH,iida-baym} 
and flux tubes \cite{iida}
in the color flavor locked (CFL) phase and in the 2SC phase \cite{Bs}. In 
these studies, superfluid vortices are topologically stable. The flux tubes 
studied so far are not topological and it is not clear if they are stable 
at all. 

All the topological string solutions like vortices or flux tubes are due to
and are related to the existence of nontrivial loops (NNL's) in the order
parameter space ($OPS$). An $OPS$ is the set of all possible values of the
order parameter ($OP$) which is the diquark condensate in this case. Each
NNL in the $OPS$ corresponds to nontrivial defect solutions in the 
superconducting phase. 
Previous studies have considered NNL's in the $OPS$ which
are generated only by the baryon number charge \cite{ZH,iida-baym}. However 
when symmetry groups such as color, flavor and baryon number are
spontaneously broken, one can have other NNL's. There are NNL's in the $OPS$
which are generated partly by the baryon number and partly by other
non-abelian color or flavor generators \cite{bal}. These loops give rise to
non-abelian strings. These string defects have been studied previously in 
the context of broken chiral symmetry in the framework of the linear sigma model
\cite{bal-digal}. In this work we consider the non-abelian defects in the
color superconducting phase. We include the effects of gauge fields which
result in flux tubes. These flux tubes are unlike those in ordinary
superconductors. The energy per unit length of the string behaves like that
of a superfluid vortex. This is why we call these non-abelian flux tubes as
semi-superfluid strings. In this work we consider the semi-superfluid
strings in the CFL phase. We will argue that these defects and superfluid
vortices are possible in the dSC, uSC and mCFL phases. In order that the
flux tubes are dynamically stable, the color superconductor must be type II. 
For asymptotic values of the chemical potential $\mu$, the color superconductor
is type I, but for intermediate values of $\mu$ it can be type II 
\cite{ren,iida-baym}.

This paper is organized as follows. In the next section, we discuss the
Ginzburg-Landau free energy and discuss the nontrivial loops in
the $OPS$ which give rise to the non-abelian string defects. Section III will
contain numerical results for the configuration of string defects in
the CFL phase. In section IV we discuss a possible scenario for the string
network and its evolution at the core of a very dense star. In
Section IV, we present our conclusions.

\section{Ginzburg-Landau free energy and the $OPS$}

In the superconducting phase the dominant pairing channel consists of 
two quarks of the same helicity. We will denote the corresponding order  
parameters by $\Phi_L$ and $\Phi_R$. $\Phi_{L,R}$ are $3\times 3$ 
matrices transforming by the $\bar{3}$ representation of $SU(3)_C$ and 
$SU(3)_{L,R}$. One can argue that a semi-superfluid string has the same 
winding number for both $\Phi_L$ and $\Phi_R$. In our calculation 
we assume that $\Phi_L=\Phi_R \equiv \Phi$. The Ginzburg-Landau(GL)
 free energy  is a function of $\Phi$. In the weak coupling and in 
the chiral limit  it is given by \cite{GR}\cite{iida-baym} 
\begin{eqnarray}\label{GL}
\Gamma&=& {1\over 4}G_{ij}^aG_{ij}^a+ {1\over 4}F_{ij}^{EM}F_{ij}^{EM} +
2\kappa_T {\rm{tr}}({\vec D}\Phi)^{\dag}({\vec D}\Phi)
+\bar{\alpha}{\rm{tr}} (\Phi^{\dag}\Phi)
+\beta_1 ( {\rm{tr}}\Phi^{\dag}\Phi)^2
+\beta_2 {\rm{tr}}(\Phi^{\dag}\Phi)^2 
\end{eqnarray}
\noindent 

If the static electromagnetic gauge field is $\vec{A}^{EM}$
and the static color gauge fields are $\vec{A}^a$, $a=1,...,8,$ the covariant
derivatives and field strengths are
\begin{eqnarray}
\vec{D}\Phi&=&{\vec\nabla}\Phi-ig{\vec{A}^aT^a} \Phi -
ie{\vec{A}^{EM}T^{EM}}\Phi,\nonumber \\
G_{ij}^a &=& \partial_i A_j^a - \partial_j A_i^a
      + g f_{abc}A_i^b A_j^c,\nonumber\\
F_{ij}^{EM} &=& \partial_i A_j^{EM} - \partial_j A_i^{EM}.
\end{eqnarray}

Now if the field is $\Phi = \Phi_L(\Phi_R)$ then the corresponding flavor
symmetry group is $SU(3)_F = SU(3)_L(SU(3)_R)$. 
Under the element $\{V_C, V_{F}, e^{i\alpha_B}\} \in SU(3)_C \times 
SU(3)_{F} \times U(1)_B$, the diquark condensate transforms as
\begin{eqnarray}\label{z3}
\Phi \to V_{F}  \Phi  V_C^{T} e^{i\alpha_B}
\end{eqnarray}
\noindent 
where $V_{F}  \in \rm{(the~\bar{3}~represenation~of)} SU(3)_{F} $, 
$V_C \in  \rm{(the~\bar{3}~represenation~of)} SU(3)_C$ and  
$e^{i\alpha_B} \in U(1)_B$.  However from Eq.(\ref{z3}), we see that the group of elements 
$\{(z_1, z_1^{-1}z_2^{-1}, z_2): z_1, z_2 = {\rm a~cube~root~of~unity}\}
 = Z_3 \times Z_3$
leaves $\Phi$ invariant. So the symmetry group of the free energy $\Gamma$
is 
\begin{equation}
G = {SU(3)_C \times SU(3)_F \times U(1)_B \over Z_3 \times Z_3}.
\end{equation}
\noindent In the symmetric or QGP phase, the diquark condensate vanishes
and so is invariant under the group $G$. On the other hand $\Phi$ is non-zero
in the superconducting phase. As a result a smaller group ($H \in G$) of
transformations keeps $\Phi$ invariant. The symmetry group $H$ depends on
the form of $\Phi$ and thereby on the state of the superconducting phase. 

In the CFL phase the free energy $\Gamma$ is minimized 
when $\Phi$ is proportional to a constant unitary matrix $U$. So one can write 
for the minimum energy configuration, $\Phi_0$, 
\begin{equation} 
\Phi_0 = \eta U
\end{equation}
\noindent with $\eta$ is a positive real number. For the analysis of the
symmetry breaking pattern one can take $\Phi_0 = \eta \bf{1}$ without loss of
generality, where $\bf{1}$ is a $3\times 3$ identity matrix. 
The group $SU(3) \times Z_3 : \{(V, V^{-1}z^{-1}, z): z = {\rm a~cube~root~of~unity}, 
V \in SU(3)\}$  keeps $\Phi_0 = \eta \bf{1}$ invariant. This set contains all 
the elements of $Z_3 \times Z_3$ defined above. In order to find the stability
group $H\subset G$ of $\Phi_0$ we must quotient this set by 
$Z_3 \times Z_3$. Hence 
\begin{equation}
H = { SU(3) \times Z_3 \over Z_3 \times Z_3}.
\end{equation}
The symmetry breaking pattern in the transition from normal to CFL phase 
($G \to H$) therefore is 
\begin{equation}
{SU(3)_C \times SU(3)_{F} \times U(1)_B \over Z_3 \times Z_3} 
\to {SU(3) \times Z_3 \over Z_3 \times Z_3}.
\end{equation}
Thus the order parameter space $OPS$ for the $\Phi$ is given by,
\begin{equation}
OPS = [SU(3)_C \times SU(3)_{F} \times U(1)_B] /
[SU(3) \times Z_3] = U(3) = {SU(3) \times U(1) \over Z_3}.
\end{equation}
\noindent
The $OPS = U(3)$ allows NNL's. In the language of homotopy groups, the
NNL's are classified by the first homotopy group $\pi_1(OPS)$. In this 
case, 
\begin{eqnarray}\label{ops}
\pi_1(OPS) = Z.
\end{eqnarray}
\noindent We now explain its features. 

$U(3)$ allows for non-abelian vortices.
We can see this as follows. A nontrivial closed loop in $U(3)$ can be
described by a curve in $SU(3)\times U(1)$ beginning at identity which becomes
closed on quotienting by $Z_3$. Consider the curve from $(\bf{1},\bf{1})$
to $(e^{i2\pi/3}, e^{-i2\pi/3})$ in $SU(3)\times U(1)$. In the $SU(3)$ part,
only the end points of this curve matter to determine the homotopy class of 
this curve. In the $U(1)$ part, it is the curve $\{e^{i\varphi} : 
0 \le \varphi \le -2\pi/3\}$
from $\bf{1}$ to $e^{-i2\pi/3}$ in the anticlockwise direction. The $SU(3)$
part of this curve is nontrivial. In $U(3)$ then, it is a nontrivial 
non-abelian closed loop $l$. It is the generator of $Z$ in Eq.(\ref{ops}). If 
$[l]$ is the homotopy class of this loop, $[l]^3$ is 
associated with a closed loop in $SU(3)$ and $U(1)$. The closed loop in 
$SU(3)$ can be deformed to a point. Hence $[l]^3$ is a NNL in 
$U(1)$ and corresponds to the abelian superfluid vortices studied in 
ref.\cite{ZH}. The elementary non-abelian vortices can be associated with 
$[l]$ or $[l]^{-1}$. Any non-abelian vortex is associated with 
$[l][l]^{3k}$ or $[l]^2[l]^{3k}$ for $k\in Z$.

Now we construct the loops in the $OPS$. We consider two loops, one in the 
homotopy class of $[l]$ and the other in the homotopy class
$[l^\prime] = [l]^{-2}$. The projections of these loops 
in the $SU(3)$ part of the $OPS$ are same. The projections of loops 
$l$ and $l^\prime$ in $U(1)_B$ go from identity to $e^{-2\pi i/3}$ in 
the anti-clockwise and clock-wise directions respectively. The latter
explicitly is $\{e^{i\varphi} : 0\le \varphi \le 4\pi/3\}$. In $U(1)_B$ 
they are generated by the baryon charge,
\begin{eqnarray}
Q_B = {2 \over 3}\bf{1}.
\end{eqnarray}
\noindent The loops $l$ and $l^\prime$ are parameterized in $U(1)_B$ as
\begin{eqnarray}
e^{-i\alpha Q_B/2},~~e^{i\alpha Q_B}
\end{eqnarray}
\noindent where the parameter $\alpha$ varies from $0$ to $2\pi$. Both
these loops start from identity and end at $e^{4\pi i/3}$ in $U(1)_B$.

Now let us discuss the projection of the loops $l$ and $l^\prime$ on
the individual groups $SU(3)_C$ and $SU(3)_F$. Note that the $U(1)_{EM}$ is
a subgroup of $SU(3)_F$. The generator $T^{EM}$ of $U(1)_{EM}$ in the
$(ds,su,ud)$ basis is given by
\begin{eqnarray}
T^{EM}={1\over 3}\left(\begin{array}{ccc}
-2 & 0 & 0 \nonumber \\
0 & 1 & 0\nonumber \\
0 & 0 & 1 \end{array} \right)\\
\end{eqnarray}
\noindent which is linear combination of the generators $S^3$ and $S^8$ of
$SU(3)_F$. So a curve generated by $T^{EM}$ lies entirely in $SU(3)_F$.
To simplify the matter we work in a basis of generators such that
$T^8(S^8)$ of $SU(3)_c(SU(3)_F)$ have the same matrix representation 
\begin{eqnarray}
{1 \over 2\sqrt{3}} \left(\begin{array}{ccc}
-2 & 0 & 0 \nonumber \\
0 & 1 & 0\nonumber \\
0 & 0 & 1 \end{array} \right)\\
\end{eqnarray}
\noindent as in ref.\cite{ren}. 
A path from $(1,1)$ to $e^{2\pi i/3}(1,1)$ in $SU(3)_C \times SU(3)_F$
can be generated by $T^{EM}$ or $T^8$ a linear combination of them. The 
projection of this curve on $SU(3)_C$ and on $SU(3)_F$ can be different. 
For example the projection of this curve can go from $\bf{1}$ to 
$e^{2\pi i/3}$ in $SU(3)_C$ and be just a point in $SU(3)_F$ or vice-versa.
Topologically there is no difference between these different
possibilities. However dynamically they are different. When the projection
in $SU(3)_C$ is from $\bf{1}$ to $e^{2\pi i/3}$ and just a point in $SU(3)_F$
the resulting string configuration will be made of only color fields with 
$g{\sqrt{3}\over 2}{\bar{\Phi}^{8}} = {2\pi}$. On the other hand if
the projection in $SU(3)_F$ is from $\bf{1}$ to $e^{2\pi i/3}$ and is just a 
point in $SU(3)_C$, the resulting string configuration will be made of only 
ordinary magnetic field with $e{\bar{\Phi}^{EM}} = {2\pi}$.
But in CFL there is a mixing between $A^{EM}$ and $A^8$ into the
following new gauge fields \cite{mixing},
\begin{eqnarray}
A_X = \cos \zeta A^{EM} + \sin \zeta  A^8 \nonumber\\
A_{\tilde{Q}} = -\sin \zeta A^{EM} + \cos \zeta  A^8
\end{eqnarray}
\noindent where $A_X$ is massive and $A_{\tilde{Q}}$ is massless. $\zeta$
depends on the couplings as follows:
\begin{equation}
\cos \zeta = \sqrt{e^2 \over e^2 + 3g^2/4}.
\end{equation}
\noindent Because of the mixing between the gauge fields a path from 
$(1,1)$ to $e^{2\pi i/3}(1,1)$ in $SU(3)_C \times SU(3)_F$ has projection
both in $SU(3)_F$ and $SU(3)_C$ for the minimum energy string configuration.
As a result for minimum energy, the sum of the ordinary and 
color magnetic flux should satisfy
\begin{eqnarray}
(e{\bar{\Phi}^{EM}}) +  g{\sqrt{3}\over 2}{\bar{\Phi}^{8}} = g_x
\bar{\Phi}^X = \pm 2\pi
\end{eqnarray}
\noindent where $\bar{\Phi}^X$ is the magnetic flux of the massive
gauge field $A^X$ and $g_x = \sqrt{e^2 + 3g^2/4}$. Note that even 
though the minimum energy configuration has a flux both of ordinary and
color magnetic fields, a configuration with only ordinary magnetic flux
can still be stable because of flux conservation.

We mention here that previous studies \cite{iida} have considered the
flux tube of the $A^X$ field. This flux tube is analogous to the electroweak
string. Unlike our case these solutions correspond to a closed loop in the
$SU(3)$ part of the $OPS$ and are not topologically stable. Their counterparts 
in the electroweak theory have been extensively studied \cite{vachaspati}
for stability. The results show that when the weak gauge coupling is
larger than the abelian gauge coupling the solutions are unstable against
expanding of the core of the string. For the same reason the flux tube
\cite{iida} considered previously in CFL will be unstable because the strong
coupling constant is an order of magnitude larger than the electromagnetic
coupling.

We expect that in the gCFL phase also there will be semi-superfluid
strings since the symmetry breaking pattern and the $OPS$ are same
as that of CFL case. In the mCFL phase we expect 
that the loops considered in the above
discussion remain non-trivial. Since different components of $\Phi$ have
different mass in this case the core structure of the defect will be
different from the case when quark masses are degenerate and are zero.
The abelian loops $[l]^{3k}$ as well as non-abelian loops $[l][l]^{3k}$ and 
$[l]^2[l]^{3k}$ still remain when quark masses are finite. Because of
spontaneous breaking of $U(1)_B$ the $OPS$ always contains a $U(1)$. 
The effect of non-zero masses only changes part of the $OPS$ which is
generated by the nonabelian generators. However this change does 
not affect the above non-abelian NNS's as long as the nonabelian generator 
of this loops $T^8$ is not explicitly broken. So there should be non-abelian
strings in the mCFL phase. The same arguments can be made about the
dSC and uSC phase. For the 2SC and g2SC cases the situation is similar to the
electroweak case. As the condensate has only one nonvanishing component, 
the loops generated by $Q_B$ and $T^8$ are same and hence 
will lie in $SU(3)_{C,F}$. So one will not have any stable topological 
string solution in the 2SC and g2SC phases.

\section{Field Equations and Numerical solutions for NA Semi-superfluid
strings}

In this section we consider the $l$ and $l^\prime$ in the $OPS$ and
derive the action and the field equation for the semi-superfluid strings.
As we mentioned above, because of the mixing between $\vec{A}^{EM}$ and
$\vec{A}_8$, the NNL's will have projections in both $SU(3)_F$ and $SU(3)_C$.
In the following, we will denote the generator corresponding to the massive 
gauge field $A_X$ by $A$. To simplify the notations we denote $T^{EM}$ by
$T$ in the following. The loops $l$ and $l^\prime$ are parameterized by

\begin{eqnarray}
M_1(\alpha)=\eta e^{i\alpha({T-Q_B/2})}
=\eta\left(\begin{array}{ccc}
e^{-i\alpha}  & 0 & 0 \nonumber \\
0 & 1 & 0\nonumber \\
0 & 0 & 1 \end{array} \right), \\
M_2(\alpha)=\eta e^{i\alpha({T + Q_B})}
=\eta\left(\begin{array}{ccc}
1  & 0 & 0 \nonumber \\
0 & e^{i\alpha} & 0\nonumber \\
0 & 0 & e^{i\alpha} \end{array} \right)
\end{eqnarray}
\noindent
with a constant $\eta$.

The projection of $\{M_1(\alpha): \alpha\in[0,2\pi]\}$ in $SU(3)$ goes from 
$I$ to the center element $e^{2\pi i/3}$ while the projection in $U(1)_{B}$
goes from $I$ to $e^{-2\pi i/3}$. The second loop
$\{ M_2(\alpha): \alpha\in[0,2\pi]\}$ is the same as $M_1(\alpha)$ except the 
path in $U(1)_{B}$ is covered with the reversed orientation.

The minimum energy configuration strings are associated with $M_1(\alpha)$ or
$M_2(\alpha)$. To keep the energy contribution from the covariant derivative to 
its minimum, only the massive gauge field $A$ is excited. 
Assuming that the string is along the $z$-axis and 
is cylindrically symmetrical around it, the $\Phi$ configuration for the string 
(corresponding to $M_1(\alpha)$ with $\alpha=\theta$, $\theta$ being the
polar angle of the position vector on the $xy$-plane) is given by
\begin{eqnarray}
\Phi(r,\theta) &=&\left(\begin{array}{ccc}
\eta f(r)e^{-in\theta}  & 0 & 0 \nonumber \\
0 & \psi_1(r) & 0\nonumber \\
0 & 0 & \psi_2(r) \end{array} \right)
 \equiv \left(\begin{array}{ccc}
\phi (r, \theta) & 0 & 0 \nonumber \\
0 & \psi_1(r) & 0\nonumber \\
0 & 0 & \psi_2(r) \end{array} \right) \\
\end{eqnarray}

To simplify our calculations we assume that $\psi_{1,2}(r)=\eta$. The string
configuration with the lowest energy will have some $r$ dependent profile
for $\psi_{1,2}(r)$. Near the core of the string $\psi_{1,2}$ will have 
values slightly larger than $\eta$ due to coupling with the $\phi$ field
and the gauge fields \cite{ren}. Far from the core of the string where
$r \to \infty$, $\psi_{1,2}(r)$ must be equal to $\eta$ but without 
any nontrivial winding like the $\phi$ field.

The finiteness of the potential energy part of the free energy 
(Eq.(\ref{GL})) requires that $f(r \to \infty) = 1$. The kinetic energy
part of the free energy can be minimized by an appropriate choice of the
gauge fields $\vec{A}$. Since the string is cylindrically symmetrical,
the phase varies only in the $\theta$ direction in the $x,y$ plane, so
only the $A_\theta$ will be non-zero. The total static free energy  
corresponding to the loop $\{ M_1(\alpha), \alpha \in [0, 2\pi] \}$ with 
$\phi=\eta f(r)e^{-in\theta}$ is given by

\begin{eqnarray}
\Gamma(\Phi,\vec{A})=
2\kappa_T |(\vec{\partial} + {i2g_x\over 3} \vec{A}) \phi|^2 +
{4g_x^2 \kappa_T \eta^2\over 9} A^2
+ \frac{1}{4} ( \partial_{\mu} A_{\nu} -
\partial_{\nu} A_{\mu}  )^2 \nonumber \\
+ {\bar \alpha} ( |\phi|^2 + 2\eta^2)
+\beta(2\phi^4+ 4\eta^2 \phi^2 + 6 \eta^4).
\end{eqnarray}
\noindent The above free energy is
minimized if, at a large distance $r$ from the string, $A_\theta$ is a
pure gauge field. This implies the following form for the gauge field
$A_\theta$:
\begin{eqnarray}
A_\theta ={\xi\gamma(r) \over r}, \gamma(r\to \infty)=1.
\end{eqnarray}
Now the covariant derivative in the $\theta$ direction is given
by
\begin{eqnarray}
D_\theta \Phi=\partial_\theta\Phi - ig_xA_\theta T\Phi
&=& \eta\left(\begin{array}{ccc}
(-i{n \over r} + {2ig_x\xi \over 3r})e^{-in\theta}
& 0 & 0 \nonumber\\
0 & - {ig_x\xi \over 3r} & 0\nonumber \\
0 & 0 & - {ig_x\xi \over 3r}\end{array} \right).\\
\end{eqnarray}

The gradient energy density from the variation of the fields along the
$\theta$ direction is given by
\begin{eqnarray}
|D_\theta\Phi|^2 = {\eta^2 \over 9 r^2}\left[(3n - 2g_x\xi)^2 +
2(g_x\xi)^2\right]
\end{eqnarray}
\noindent which is minimized by $g_x\xi = 1 $ for $n=1$. So the
gauge field $A_\theta$ takes the form
\begin{eqnarray}
A_\theta ={\gamma(r) \over g_xr}.
\end{eqnarray}

It is important to note that the gauge field reduces the gradient
energy by ${1\over 3}$ for the string configuration corresponding to loop
$M_1(\theta)$. 

The Euler-Lagrange equations for $\Phi$ and the gauge field $A$ respectively
are 
\begin{eqnarray}
\partial_{\mu} \frac{\delta \Gamma}{\delta (\partial \phi ^{*})}
-\frac{\delta \Gamma}{\delta \phi^{*}}
&=& 2 \kappa_T
  (\partial _{\mu} + {2ig_x\over 3} A_{\mu} )
  (\partial _{\mu} + {2ig_x\over 3} A_{\mu} ) \phi
- (\bar \alpha + 4 \beta\eta^2) \phi - 4 \beta |\phi|^2 \phi=0,\\
\partial_{\mu} \frac{\delta \Gamma}{\delta (\partial A_{\mu})}
-\frac{\delta \Gamma}{\delta A_{\mu}}
&=& \partial ^2 A_{\mu}
+ {4 \kappa_T ig_x\over3} (\phi^{*} \partial_{\mu} \phi
- \phi \partial_{\mu} \phi^{*})
- {16 \kappa_T g_x^2\over 9} |\phi|^2 A_\mu
- {8 \kappa_T g_x^2\eta^2\over 9}A_\mu =0.
\end{eqnarray}

The static equations satisfied by $f$ and $\gamma$ are
\begin{eqnarray}
&&
f^{''}(r) +\frac{1}{r} f^{'}(r) - \frac{f(r)}{r^2}
\left({4\gamma(r) \over 3} -1\right)^2
-\frac{\bar{\alpha}+ 4 \beta \eta^2}{ 2 \kappa_T} f(r)
-\frac{2 \beta}{\kappa_T} \eta^2 f^3(r) =0,  \label{diff1} \\
&&\gamma^{''}(r) -\frac{1}{r} \gamma^{'}(r)
-8\kappa_Tg_x^2 \eta^2\left({2f^2(r)+ 1\over 9}\right) \gamma(r)
+ {8 \kappa_T g_x^2 \eta^2 f^2(r)
\over 3}=0.   \label{diff2}
\end{eqnarray}

Now let us consider the string solution for the loop given by $M_2(\theta)$.
The ansatz for the $\Phi$ field in this case is
\begin{eqnarray}
\Phi(r,\theta)&=&\left(\begin{array}{ccc}
\psi(r) & 0 & 0 \nonumber \\
0 & \eta f(r)e^{in\theta} & 0\nonumber \\
0 & 0 & \eta f(r)e^{in\theta} \end{array} \right). \\
\end{eqnarray}

As in the previous case of the string solution corresponding to the loop 
$M_1$, we assume that $\psi(r)=\eta$ to simplify our calculations.
The free energy with $\phi=\eta f(r)e^{in\theta}$ is given by

\begin{eqnarray}
\Gamma( \Phi, \vec{A})=
4\kappa_T |(\vec{\partial} - {i g_x \over 3}\vec{A}) \phi|^2
+ {8g_x^2 \kappa_T\eta^2\over 9} A^2
+ \frac{1}{4} ( \partial_{\mu} A_{\nu} -
\partial_{\nu} A_{\mu})^2 \nonumber \\
+ {\bar \alpha} (2|\phi|^2 + \eta^2)
+\beta(6\phi^4+ 4\eta^2 \phi^2 + 2 \eta^4).
\end{eqnarray}

Again since the phase of the condensate depends only on $\theta$, only
$A_\theta$ is non-zero. To find out the asymptotic form of $A_\theta$, 
we impose the condition that it asymptotically becomes a pure gauge field and also minimizes the 
gradient energy. The covariant derivative of $\Phi$ is

\begin{eqnarray}
D_\theta \Phi=\partial_\theta\Phi - ig_xA_\theta T\Phi
&=& \eta\left(\begin{array}{ccc}
{2ig_x\xi \over 3r} & 0 & 0 \nonumber\\
0 & (i{n \over r}- {ig_x\xi \over 3r})e^{in\theta} & 0\nonumber \\
0 & 0 & (i{n \over r}- {ig_x\xi \over 3r})e^{in\theta}
\end{array} \right).\\
\end{eqnarray}

The gradient energy density corresponding to the phase variation of $\Phi$
is now given by

\begin{eqnarray}
|D_\theta\Phi|^2 = {1 \over 9 r^2}\left[2(3n - g_x\xi)^2 +
4(g_x\xi)^2\right]
\end{eqnarray}
\noindent which is again minimized by $g_x\xi=1$. The reduction in the gradient 
energy due to the gauge fields is now only by a factor of ${2\over 3}$. The
Euler-Lagrange equations for the fields $\phi$ and $\vec{A}$ are

\begin{eqnarray}
\partial_{\mu} \frac{\delta \Gamma}{\delta (\partial \phi ^{*})}
-\frac{\delta \Gamma}{\delta \phi^{*}}
&=& 4 \kappa_T
  (\partial _{\mu} - {ig_x\over 3} A_{\mu} )
  (\partial _{\mu} - {ig_x\over 3} A_{\mu} ) \phi
-2 (\bar \alpha + 2\beta\eta^2) \phi - 12 \beta |\phi|^2 \phi=0,\\
\partial_{\mu} \frac{\delta \Gamma}{\delta (\partial A_{\mu})}
-\frac{\delta \Gamma}{\delta A_{\mu}}
&=& \partial ^2 A_{\mu}
- {4 \kappa_T ig_x\over3} (\phi^{*} \partial_{\mu} \phi
- \phi \partial_{\mu} \phi^{*})
- {8 \kappa_T g_x^2\over 9} |\phi|^2 A_\mu
- {16 g_x^2\eta^2\over 9}A_\mu =0
\end{eqnarray}
\noindent which simplifies to the following:
\begin{eqnarray}
&&
f^{''}(r) +\frac{1}{r} f^{'}(r) - \frac{f(r)}{r^2}
\left({\gamma(r) \over 3} + 1\right)^2
-\frac{\bar{\alpha}+ 2\beta \eta^2}{ 2 \kappa_T} f(r)
-\frac{3 \beta}{\kappa_T} \eta^2 f^3(r) =0,  \label{diff3}\\
&&\gamma^{''}(r) -\frac{1}{r} \gamma^{'}(r)
-8\kappa_Tg_x^2 \eta^2 \left({f^2(r)+ 2 \over 9}\right) \gamma(r)
+ {8 \kappa_T g_x^2 \eta^2 f^2(r)
\over 3}=0.\label{diff4}
\end{eqnarray}

We solve equations (\ref{diff1})-(\ref{diff2}) and (\ref{diff3})-(\ref{diff4}) numerically to find out the string
profile for the two loops $l$ and $l^\prime$ respectively. We choose the values of parameters which will correspond to a type II color superconductor.
For our calculations we took $g_x = 2.0$, $\kappa_T=0.42$, 
$\beta = 1.26$ and $\eta=100$MeV.  The parameter $\bar{\alpha}$ is obtained 
from the relation $\bar{\alpha} = 8 \beta \eta^2 = 1.008 \times 10^5$ MeV$^2$. 
We also use the following relations \cite{taeko}:
\begin{eqnarray}
\frac{ \bar{\alpha} }{\kappa_T}
\sim \frac{\beta \eta^2}{\kappa_T} \sim m_{\phi}^2,\nonumber \\
2 \kappa_T g_x^2 \eta^2 \equiv m_{A}^2.  \nonumber \\
\end{eqnarray}
For these parameters we have the Higgs mass $m_\Phi \sim 245$MeV and the 
Meissner mass (inverse of the penetration depth) $m_A \sim 183$MeV.

The results of our numerical solutions are shown in Fig.1 and Fig.2. Though 
the figures look similar, the energy of the configurations are very different.
One can see that both the $\phi$ and $A$ profiles vary more slowly and reach 
their asymptotic values at larger $r$ for the string corresponding to the NNL 
$l^{\prime}$. The NNL $l^\prime$ in the $OPS$ travels a longer path in $U(1)_B$ 
which costs larger gradient energy for the semi-superfluid solutions as 
$U(1)_B$ is a global symmetry group.
\begin{figure}[hbt]
\begin{center}
\includegraphics[width=12cm]{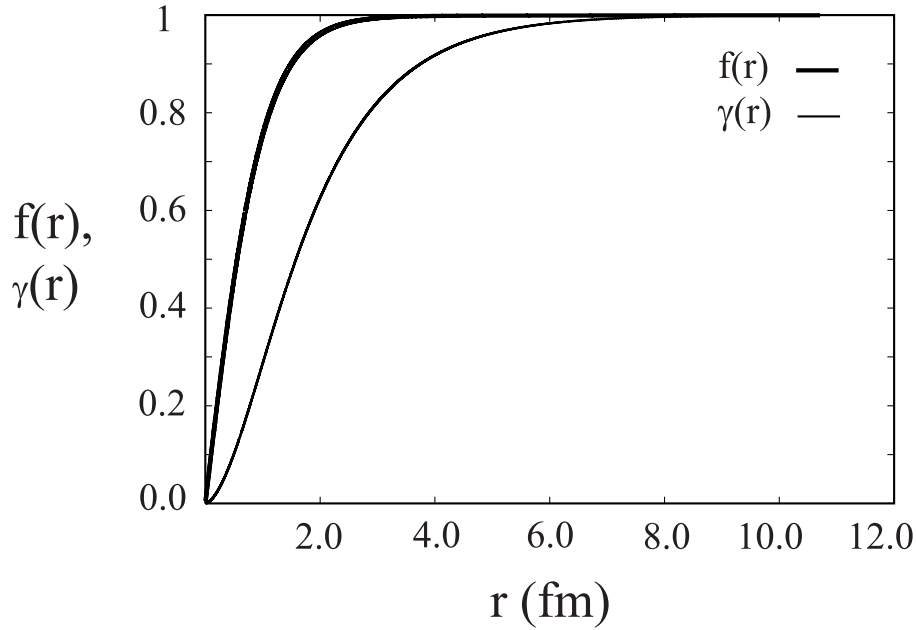}
\end{center}
\caption{The $f(r)$ and $\gamma(r)$ profiles for the string corresponding
to the non-trivial loop $l$.}
\end{figure}

\begin{figure}[hbt]
\begin{center}
\hskip-3cm
\includegraphics[width=12cm]{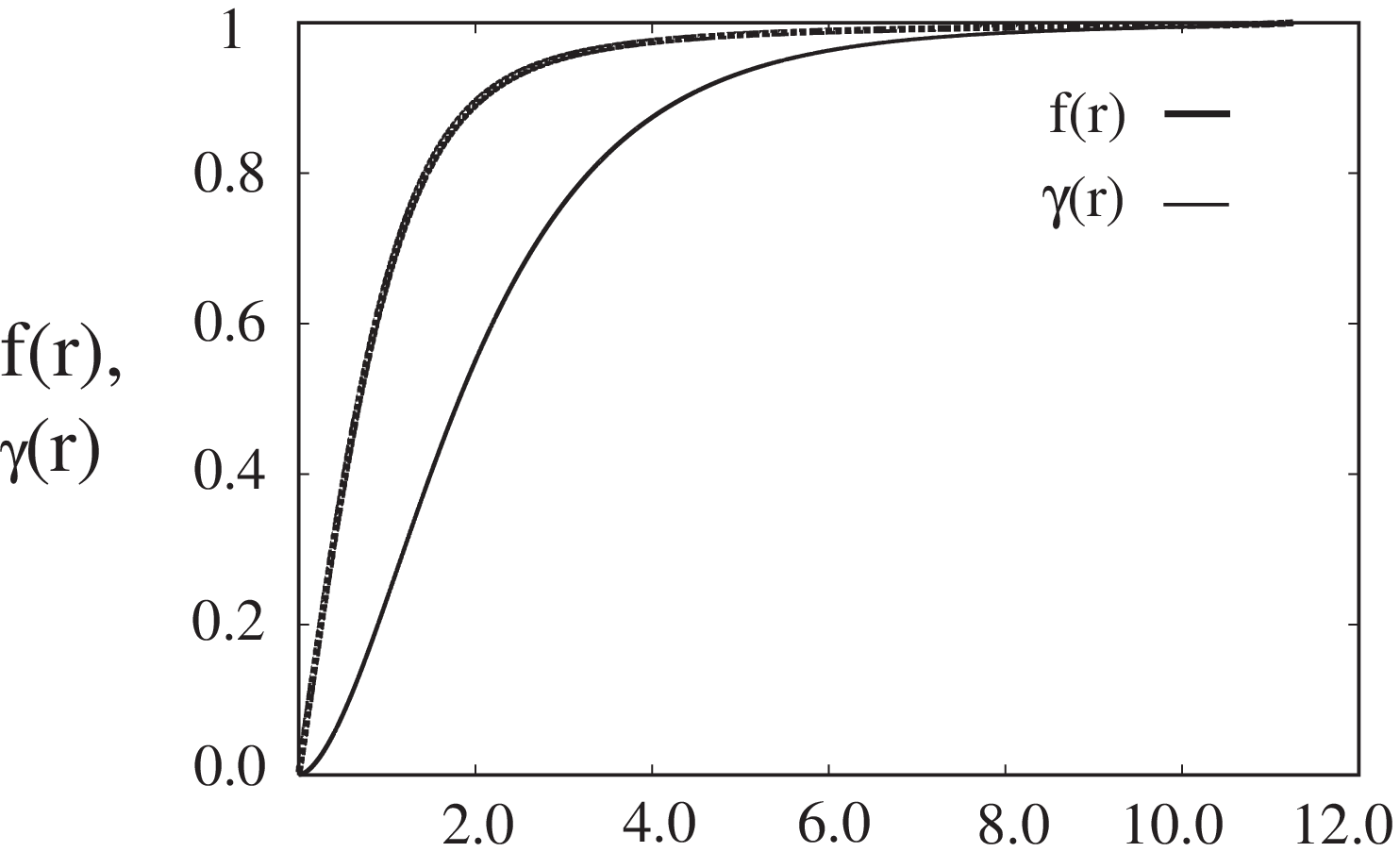}
\end{center}
\caption{$f(r)$ and $\gamma(r)$ profile for string corresponding
to the non-trivial loop $l^\prime$}
\end{figure}
In the profiles of the string, we see that far from the core of
the semi-superfluid string, the field is color-flavor locked, i.e. the
field is a constant times a $SU(3)$ matrix. However at the core of the
string the condensate is not locked.

The NA string corresponding to the loop $M_1(\alpha)^3$ and a winding number
one superfluid string are topologically equivalent. For both these string
configurations the energy density is $\sim{3\over r^2}$ at large $r$.
It is not clear if one of these configuration decays into the other. On the
other hand, this NA string is also topologically equivalent to three elementary
NA strings. However we do not believe that three clearly separated NA elementary
strings will evolve into the above mentioned NA string or the superfluid
string. We expect that elementary superfluid strings of the same winding 
number repel each other at large separations. So the total flux of a network of
clearly separated strings should be additive. We address this issue in a
future work.

\section{Formation and Implication of Strings in CFL phase}

Our non-abelian string solutions are the only stable topological string 
solutions with quantized flux of ordinary as well as color magnetic field.
All other solutions with flux tube are topologically unstable for realistic 
values of the strong and electromagnetic couplings. It is interesting to note
that when one takes an electron in a closed path around the non-abelian
strings, the Aharonov-Bohm phase is different from $2\pi$. This will lead to
strong scattering of electrons from these strings. In the following we
discuss a possible scenario for the formation of string defects inside the
core of very dense stars.

Semi-superfluid or superfluid defects in the CFL phase can form during
the phase transition from QGP to CFL phase. Topological strings can also be 
induced in the
CFL phase from the outer surrounding confining medium when the star starts
to spin up. This is in analogy with creation of vortices in rotating
superfluid. In the following, we discuss the formation of strings during the
QGP-CFL transition inside very dense stars.

A phase transition from the QGP to CFL phase may be expected inside the core
of dense stars while the star is cooling. Topological strings as well as
non-topological strings will be formed during this transition. We assume that
this transition is of first order. In this case, the transition takes place via
nucleation of CFL bubbles in the QGP background. When the temperature inside
the star cools below the critical temperature $T_c$, bubbles of CFL phase
nucleate in the QGP background. Inside the bubbles, the magnitude of the
condensate is uniform and the massive gauge field $A$ is zero. 
The phase $\alpha$ of the condensate is also uniform inside 
the bubble and varies randomly from
one bubble to the other. When the bubbles meet, $\alpha$ interpolates between
the values of the phase in the two colliding bubbles. There are two possible
ways $\alpha$ can interpolate in the presence of gauge fields \cite{ajit}.
Numerical experiments have shown that the interpolation of $\alpha$ is such
that the total variation of $\alpha$ is the lowest \cite{rhb}. When 
$\alpha$ starts to interpolate, the massive gauge field $A$ also gets excited 
to minimize the gradient energy. When three or more bubbles collide, it 
sometimes leads to the variation of $\alpha$ by an integer multiple of
$2\pi$ around the loop at the intersection point. When this happens, a
semi-superfluid string is formed. This is the conventional mechanism of
defect formation known as the Kibble mechanism \cite{kibble}. However there
are other mechanisms which also contribute to the defect density 
\cite{dgl1}.

One may expect that the presence of strong external magnetic fields will
affect the formation of semi-superfluid strings. However since the unbroken
gauge field in the CFL phase consists $\sim 99\%$ of the electromagnetic
field, the major part of the external magnetic field will propagate
unscreened \cite{alford}. Only a very small fraction of the external magnetic
field, basically related to the massive gauge field, will be repelled by 
the CFL phase. So in a sense the formation of semi-superfluid strings is more
like the formation of flux tubes in ordinary superconductors under a small
external field. Note that the massive gauge field is made up mostly
of the color gauge field, so our flux tube will consist mainly of color 
magnetic flux. We expect that there is no long range color magnetic field
in the medium before the transition takes place. So the formation of
semi-superfluid strings during the transition is spontaneously induced rather 
than external field-induced.

Now we discuss a possible scenario for the network of strings inside the
CFL core of the dense star. Usually the temperature at the center of the
star is higher and gradually decreases as one moves radially outwards.
So when the star cools, the transition from the quark-gluon plasma (QGP) 
to the CFL phase will take 
place first in a thin spherical shell where the temperature drops below 
the critical temperature. In this thin spherical region bubbles of CFL
nucleate and grow. As they grow the bubbles coalesce with other bubbles.
However since the temperature is higher towards the center of the star
the bubbles will grow mostly in the spherical region forming a thin
spherical shell of CFL. The picture of the phase distribution is that of 
a spherical shell of the CFL phase covering the QGP core with temperature 
above $T_c$. The QGP and CFL phases are separated by the QGP-CFL boundary. 
The CFL shell and the outer confined crust are separated by the 
CFL-confining boundary.
At this point, the cooling of the star will be different from that of cooling 
due to neutrino emission because there will be generation of latent heat 
when QGP converts into CFL.

Further dynamics of the transition can be either by bubble nucleation or
motion of the interior wall of the spherical CFL shell covering QGP. However
these two pictures are not much different because even if there is bubble
nucleation, the bubbles will nucleate close to the boundary wall as the
temperature farther inside is either close to $T_c$ or higher. So the basic
picture of transition is that a CFL shell is formed due to bubble nucleation
and then the phase transition takes place through the motion of the interior
wall of the CFL shell towards the center of the star.

The strings which can survive such a transition are those which are oriented
along the radial direction. One end of this string will end on the inner
QGP-CFL boundary and the other in the CFL-confining boundary. The strings
can end on the CFL-confining boundary because of the availability of
color monopole and anti-monopoles pairs in the confined phase \cite{alford}.
The strings with both ends connected to the outer confining crust will decay
by shrinking to the crust. The surviving radially oriented strings will
possibly increase in length along the radial direction. However the density
of string ends on the 
inner CFL-QGP boundary will increase due to the shrinking QGP
core. Eventually the density will be so high that ends of different strings
will come in contact with each other. One can argue that the total number of
strings ending on the QGP-CFL boundary are even in number with equal number
of strings and anti-strings. Two such strings with opposite winding will join 
together forming huge $V$-shaped strings with ends connected to the outer 
confining crust. These $V$-shaped strings are unstable to moving towards the 
outer confining crust. This movement happens almost simultaneously
to all string-antistrings pairs ending on the QGP-CFL boundary. Such an
evolution of a network of strings may affect the properties of a star
like its angular momentum.

\section{Conclusions}

We have studied non-abelian semi-superfluid strings in 
high density QCD. Inclusion
of gauge fields reduces the energy of these strings compared with the
$U(1)_B$ superfluid string. Even with the gauge fields the energy per unit
length of the semi-superfluid string is logarithmically divergent with
the system size. Still such strings are relevant for finite system like stars. 
This is unlike the flux tubes in ordinary superconductors. 
The semi-superfluid strings are partly like superfluid strings and partly like flux tubes. 
These are the only topological
strings possible in high density QCD which have flux of ordinary magnetic and
color magnetic fields. Parallel transport of an electron in a closed loop
around these strings picks up an Aharonov-Bohm phase different from $2\pi$
leading to their strong scattering from the semi-superfluid string. We
propose a scenario for a string network at the core of the star, the 
evolution of which can affect the dynamics of the star. However a detailed
study using realistic phase transition dynamics for the QGP-CFL transition
and formation of strings is necessary to make any definite prediction for
the string network or their possible effects. We plan to do such calculations
in the future.

\begin{acknowledgments}
We are grateful to T. Hatsuda, M. Alford, I. Giannakis, H-c. Ren for  
helpful comments and discussions. We also like to thank K. Fukushima, 
K. Iida, R. Ray and A. M. Srivastava for useful discussions.
S.D. and T.M. are supported by the Japan Society for the Promotion of Science
for Young Scientists. This work was also supported by DOE under contract 
number DE-FG02-85ER40231.
\end{acknowledgments}

\end{document}